\documentclass[10pt]{iopart}
\expandafter\let\csname equation*\endcsname\relax

\expandafter\let\csname endequation*\endcsname\relax

\usepackage{graphicx}%
\usepackage{multirow}%
\usepackage{amsmath,amssymb,amsfonts}%
\usepackage{amsthm}%
\usepackage{mathrsfs}%
\usepackage{xcolor}%
\usepackage{textcomp}%
\usepackage{subcaption}
\usepackage{cuted}

\usepackage[noadjust,compress]{cite}
\usepackage{lipsum}

\usepackage[colorlinks=true,linkcolor=blue, citecolor=blue, urlcolor=blue]{hyperref}
\begin{document}
\title[Kerr-Newman black hole with quintessence and a cloud of string.]{Thermodynamics and null-geodesic of the Kerr-Newman black hole surrounded by quintessence and a cloud of string.}

\author{Aheibam Boycha Meitei$ ^{1} $ \footnote{ E-mail: aheibamboycha143@gmail.com}, Yenshembam Priyobarta Singh$^{2}$ \footnote{ E-mail: priyoyensh@gmail.com}, Telem Ibungochouba Singh$^{2,*}$ \footnote{$^*$
 E-mail: ibungochouba@rediffmail.com (corresponding author) }, 
Irom Ablu Meitei$^{1}$ \footnote{ E-mail: ablu.irom@gmail.com}, Kangujam Yugindro Singh$ ^{1} $ \footnote{E-mail: yugindro361@gmail.com}}
\address{$^1$ Department of Physics, Manipur University, Canchipur, Imphal 795003, Manipur, India.}
\address{$^2$ Department of Mathematics, Manipur University, Canchipur, Imphal 795003, Manipur, India.}


\begin{abstract}
In this paper, we study the effect of the modified dispersion relation (MDR) on the thermodynamics of the Kerr-Newman black hole surrounded by quintessence and a cloud of string. The thermodynamic properties of the Kerr-Newman black hole are shown to rely not only on the black hole's properties but also on the parameters associated with the modified dispersion relation, quintessence, and the cloud string. Additionally, the equation of state is impacted by these parameters. The remnant and the stability of the black hole are also discussed under the correction of MDR, quintessence, and a cloud of string. In addition, the null geodesic structure of the spacetime is studied using the Hamilton--Jacobi formalism. The effective potential for photon motion is obtained, and the radii of the prograde and retrograde circular photon orbits are determined. The instability of these circular photon orbits is further characterized by the Lyapunov exponent.\\


\end{abstract}
\maketitle

\ioptwocol
\section{Introduction}

Hawking's black hole area theorem \cite{1,2,3} from the early 1970s, along with Bekenstein's \cite{4,5,6} view of the event horizon's surface area as an indicator of a black hole's entropy, signifies the onset of black hole thermodynamics. In classical physics entering a black hole is a one-way process. However, Stephen Hawking, using the ideas of quantum field theory in curved spacetime, showed that black holes emit particles and the energy spectrum of the emitted particles is thermal \cite{7,8}. The fundamental concept of black hole thermodynamics suggests that the irreversibility experienced when an object is absorbed by a black hole resembles the statistical irreversibility commonly observed in regular physics \cite{9}. Black hole thermodynamics illustrates the profound relationship between general relativity and quantum theory, highlighting the quantum nature of the black holes.

Lorentz symmetry is a principle that mandates the laws of physics remain consistent for all observers \cite{10}. It is a cornerstone of Einstein's special relativity and is essential in quantum field theory, particle physics, and cosmology. This symmetry has been observed to break down at extremely high energy levels, specifically at the Planck energy scale due to the discreteness of space-time in the Planck scale \cite{11}, spontaneous symmetry breaking \cite{12},or a vector field constrained kinematically to be timelike called aether-like field \cite{13,13a,13b,13c,13d,13e}. Planck energy acts as a boundary distinguishing classical physics from quantum physics. The violation of Lorentz symmetry can be expressed in the form of modified dispersion relations (MDRs). The modified energy-momentum dispersion relationship emerges from a new theory referred to as doubly special relativity (DSR) which is an extension of the special theory of relativity \cite{14,15,16}. In DSR, there are two constant quantities: one is the Planck energy and the other is the speed of light. The impact of MDR on the thermal quantities and the evaporation of black hole is discussed in refs \cite{17,18,19,20,21,22,23,24}.

With the discovery of dark energy, it has become clear that our universe is not matter-dominated, and most of the energy in the universe is gravitationally self-repulsive, causing the expansion of the universe to accelerate \cite{25,26,27,28,29,30}. Different dark energy models are proposed namely the cosmological constant, quintessence energy etc. Currently, the cosmological constant is understood as an energy inherent to the vacuum of space that possesses negative pressure and drives cosmic acceleration. It maintains a uniform value throughout the entirety of space across all time and is chemically non-reactive \cite{31}. Now, we employ quintessence rather than the cosmological constant, even though their impacts on the Universe's expansion are comparable. This is due to the fact that the experimentally determined value of the cosmological constant is significantly less than its anticipated theoretical value \cite{32}. Quintessence is characterized by the equation of state $\omega = \frac{p}{\rho}$, in which $ p $ represents pressure and $ \rho $ signifies the energy density, and it falls within the interval $ -1<\omega<-\frac{1}{3} $. Kiselev \cite{33} integrated black hole physics with quintessence matter by introducing a new static, spherically symmetric exact solution to Einstein's equations. Research is starting to examine how quintessence affects various types of black holes. Chen et al. \cite{34} studied the thermodynamics and present a solution of Einstein's equation with quintessence matter surrounding a d-dimensional black hole. Wei et al. \cite{35,36} also investigated the impact of quintessence on the thermal quantities of the Reissner-Nordstrom black hole. Hamil et al. \cite{37,38,39,40} studied the thermodynamics of the schwarzschild and Reissner-Nordström black hole surrounded by quintessence. There are many other papers working under the background of the quintessence \cite{41,42,43,44,45,45a,45b}.

The cloud of strings model, proposed by Letelier \cite{46,47,48}, provides a specific structure for characterizing spacetime shaped by a continuous array of strings. This model is one of the important aspect in gravitational physics. Incorporating cloud of string model into black hole solutions modify the geometry by adding an extra parameter $ b $, which modifies the black hole mass and the structures of the horizon \cite{49}. Numerous researchers have started investigating how the cloud of string model affects different black hole solutions. Toledo et al. \cite{50} investigated the solution related to the Reissner–Nordström black hole that is surrounded by quintessence and a cloud of string. Costa et al. \cite{51} also derived the solution for a static, spherically symmetric black hole surrounded by a cloud of strings within a quintessential fluid. Cárdenas et al. \cite{52} expressed their apprehensions regarding the implementation of general relativistic tests on the spacetime generated by a static black hole linked to a cloud of strings in a universe that contains quintessence. Mustafa et al. \cite{53} investigated the motion of null and timelike geodesics in the vacinity of the Schwarzschild black hole, considering the effects of both the cloud of string and the quintessence field. In various studies, many researchers have explored black hole solutions that incorporate a quintessence field, a string cloud, or both in various black hole configurations \cite{54,55,56,57,58,59,60,61,62,63}. In this paper, considering the MDR, quintessence, and a cloud of string, we investigate the thermodynamic properties and the null-geodesic equations of the Kerr-Newman black hole and the role played by each component.

The paper is organised as follows. In section 2, we introduced a brief review of the MDR. In section 3, Kerr-Newman black hole surrounded by quintessence and a cloud of string is discussed. In section 4, we investigate the effects of the MDR, quintessence, and a cloud of string on the thermodynamic properties of the Kerr-Newman black hole. In section 5, null geodesic equations of the Kerr-Newman black hole surrounded by quintessence and cloud of string are derived. In section 6, the Lyapunov exponent and instability of circular photon orbit are discussed. Discussion and conclusion are given in section 7.

In this paper, we use the metric signature $ (-,+,+,+) $ and the natural unit system $( \hbar= c=G=1 ) $.

\section{A brief review of the MDR}  
\renewcommand {\theequation}{\arabic{equation}}
Quantum gravity theories, including string theory and loop quantum gravity, suggest that energy-momentum dispersion relation is violated at extremely high energy levels, specifically at the Planck scale. There are two possible ways of modification of energy-momentum dispersion relation \cite{64}. 

The first one is with Planck-scale correction of order $ l_{p}E^{3} $ as
\begin{eqnarray}\label{1}
p^{2}=E^{2}-\mu^{2}+\eta_{1}l_{p}E^{3}.
\end{eqnarray}
The second one is with Planck-scale correction of order $ l^{2}_{p}E^{4} $
\begin{eqnarray}\label{2}
p^{2}=E^{2}-\mu^{2}+\eta_{2}l^{2}_{p}E^{4}.
\end{eqnarray}
Here, $ \eta_{1} $ and $ \eta_{2} $ are the quantum correction parameters and $ \mu $ is the mass parameter which is proportional to the rest energy.

In order to determine the variation in the positions of the particles and the uncertainty in their energy resulting from the modified dispersion relation, we apply a variation to Eq. \eqref{1} and Eq. \eqref{2} as
\begin{eqnarray}\label{eq.3}
\delta p \simeq \left(1+\eta_{1} l_{p}E \right) \delta E
\end{eqnarray}
and
\begin{eqnarray}\label{eq.4}
\delta p \simeq \left(1+\frac{3}{2}\eta_{2} l^{2}_{p}E^{2} \right) \delta E.
\end{eqnarray}

By setting $ \mu=0 $ and presuming $ E\simeq \delta E $, we used the standard Heisenberg uncertainty principle, $ \delta E\geq \frac{1}{\delta x} $and $ \delta p\geq \frac{1}{\delta x} $ in Eqs. \eqref{eq.3} and \eqref{eq.4}, we get
\begin{eqnarray}\label{eq.5}
E\geq \frac{1}{\delta x}\left(1-\frac{\eta_{1}l_{p}}{\delta x} \right)
\end{eqnarray}
and
\begin{eqnarray}\label{eq.6}
E\geq \frac{1}{\delta x}\left(1-\frac{3\eta_{2}l^{2}_{p}}{2(\delta x)^{2}} \right).
\end{eqnarray}

\section{Kerr-Newman black hole surrounded by quintessence and a cloud of string}
The metric associated with a rotating black hole surrounded by quintessence and a cloud of strings is given by \cite{65}
\begin{eqnarray}\label{eq.7}
ds^{2}&=&-\frac{\Delta -a^{2}\sin^{2}\theta}{\Sigma}dt^{2}+\frac{\Sigma}{\Delta}dr^{2}-2a \sin^{2}\theta \cr&& \times\left(1-\frac{\Delta -a^{2}\sin^{2}\theta}{\Sigma} \right)dt d\varphi +\Sigma d\theta ^{2} + \sin^{2}\theta \cr&& \times\left[\Sigma + a^{2}\sin^{2}\theta\left(2-\frac{\Delta -a^{2}\sin^{2}\theta}{\Sigma} \right)  \right]d\varphi ^{2} ,
\end{eqnarray}
where
\begin{eqnarray}
\Sigma &=& r^{2}+a^{2}cos^{2}\theta ,\cr
\Delta &=&(1-b)r^{2}+a^{2}+Q^{2}-2Mr-\alpha r^{-3\omega +1}.
\end{eqnarray}
Here, $M,a,Q,b$ and $\alpha$ represent the black hole mass, spin parameter, charge, cloud string parameter, and the quintessence parameter of the Kerr-Newman black hole.

Due to the rotation of the Kerr-Newman black hole, it induces frame-dragging in the surrounding area and compels particles and photons within the ergosphere to accelerate in the direction of the black hole's rotation at an angular velocity $\Omega=\frac{g_{03}}{g_{33}}$.

Performing the coordinate transformation as \cite{66,67} 
\begin{eqnarray}
d\phi &=& d\varphi-\Omega dt\cr &=& d\varphi -\frac{a \Pi}{(r^{2}+a^{2})\Sigma^{2}+a^{2}\Pi sin^{2}\theta}dt,
\end{eqnarray}
where $ \Pi =(2Mr-Q^{2}+\alpha r^{1-3\omega}+br^{2}) $.
Eq. \eqref{eq.7} becomes
\begin{eqnarray}
ds^{2}&=&  - \frac{\Delta}{r^{2}+a^{2}+a^{2}\sin^{2}\theta\left( \frac{2Mr-Q^{2}+\alpha r^{1-3\omega}+br^{2}}{\Sigma^{2}}\right)} dt^{2}\cr&& + \frac{\Sigma}{\Delta}dr^{2} +\Sigma^{2}d\theta^{2}++ \sin^{2}\theta \cr&& \times \left[\Sigma + a^{2}\sin^{2}\theta\left(2-\frac{\Delta -a^{2}\sin^{2}\theta}{\Sigma} \right)  \right] d\phi ^{2}.
\end{eqnarray}

The surface gravity at the event horizon $(r_{h})  $ of a Kerr-Newman black hole, surrounded by quintessence and a cloud of string, is defined as \cite{45}
\begin{eqnarray}
\kappa &=& \lim_{g_{00}\to 0} \left( -\frac{1}{2}\sqrt{-\frac{g^{11}}{g_{00}}}\frac{dg_{00}}{dr}\right),\cr &=& \frac{r_{h}-M-br_{h}-\frac{r_{h}^{-3\omega}\alpha (1-3\omega)}{2}}{r_{h}^{2}+a^{2}}.
\end{eqnarray}
By setting the equation $\Delta = 0$, the boundary of the black hole can be determined, allowing us to define the mass of the black hole as
\begin{eqnarray}\label{eq.12}
M=\frac{1}{2}\left[(1-b)r_{h}+\frac{a^{2}}{r_{h}}+\frac{Q^{2}}{r_{h}}-\alpha r_{h}^{-3\omega} \right]. 
\end{eqnarray}

\begin{figure}[h!]
\centering

\begin{subfigure}{\columnwidth}
\includegraphics[width=\linewidth]{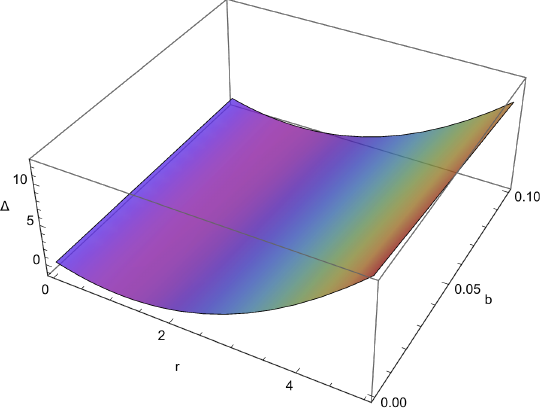}
\caption{}
\end{subfigure}

\begin{subfigure}{\columnwidth}
\includegraphics[width=\linewidth]{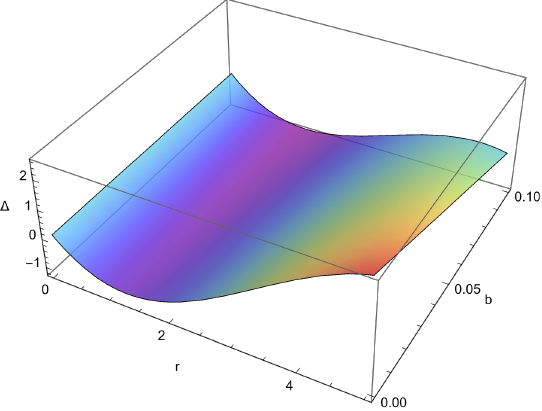}
\caption{}
\end{subfigure}

\begin{subfigure}{\columnwidth}
\includegraphics[width=\linewidth]{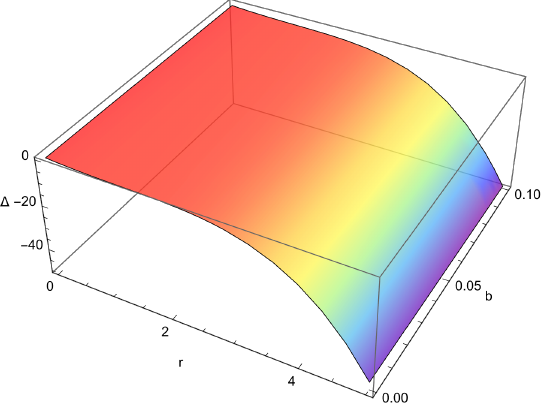}
\caption{}
\end{subfigure}
\label{fig Func}
\caption{Variation of the function $ \Delta $ with respect to $ r $ and $ b $ for (a) $ \omega=-\frac{1}{3} $ (b) $ \omega=-\frac{2}{3} $ and (c) $ \omega=-1 $. Here, we used $a=0.1$  , $ \alpha=0.1 $, $ Q=0.3 $ and $ M=1 $.}
\end{figure}
\begin{figure}[h!]
\centering

\begin{subfigure}{\columnwidth}
\centering
\includegraphics[width=\linewidth]{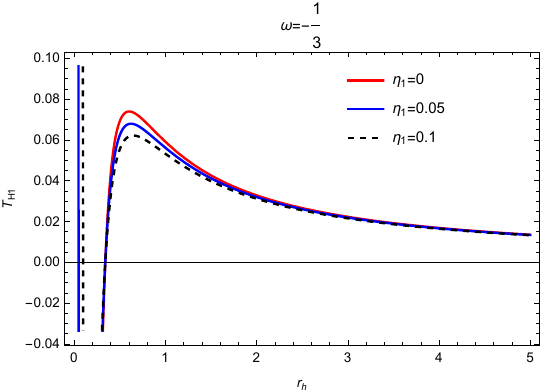}
\caption{}
\label{fig:graph1}
\end{subfigure}

\vspace{0.3cm}

\begin{subfigure}{\columnwidth}
\centering
\includegraphics[width=\linewidth]{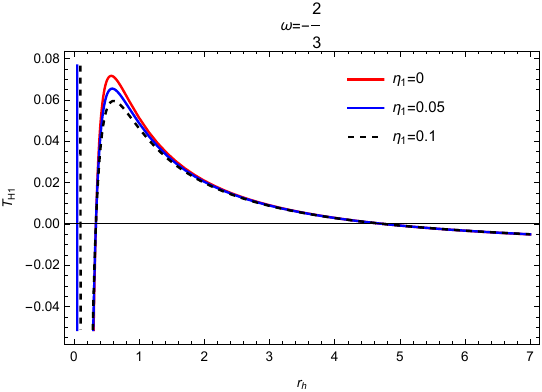}
\caption{}
\label{fig:graph2}
\end{subfigure}

\vspace{0.3cm}

\begin{subfigure}{\columnwidth}
\centering
\includegraphics[width=\linewidth]{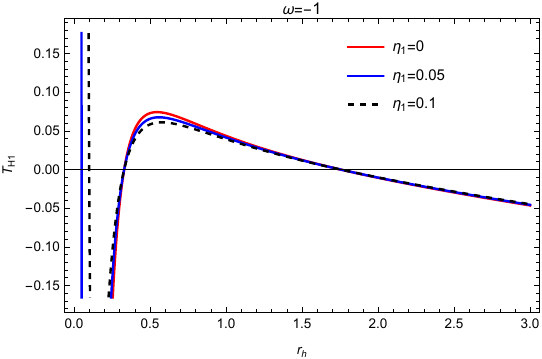}
\caption{}
\label{fig:graph3}
\end{subfigure}

\caption{Temperature ($ T_{H1} $) vs radius of event horizon ($ r_{h} $) graph for $a=0.1$  , $ \alpha=0.1 $, $ b=0.05 $, $ Q=0.3 $ and $ l_{p}=1 $.}
\label{Tem1}

\end{figure}
\section{Thermodynamics of the Kerr-Newman black hole surrounded by quintessence and cloud of string with MDR}
\begin{figure}[h!]
\centering

\begin{subfigure}{\columnwidth}
\centering
\includegraphics[width=\linewidth]{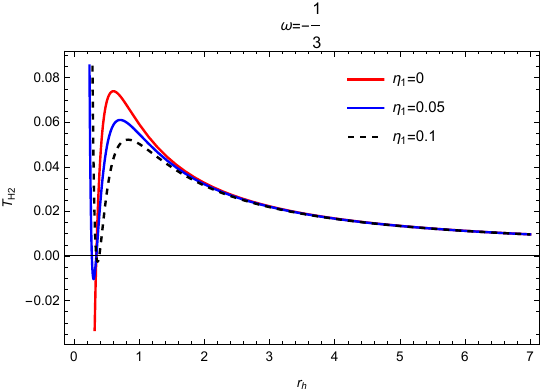}
\caption{}
\label{fig:graph1}
\end{subfigure}

\vspace{0.3cm}

\begin{subfigure}{\columnwidth}
\centering
\includegraphics[width=\linewidth]{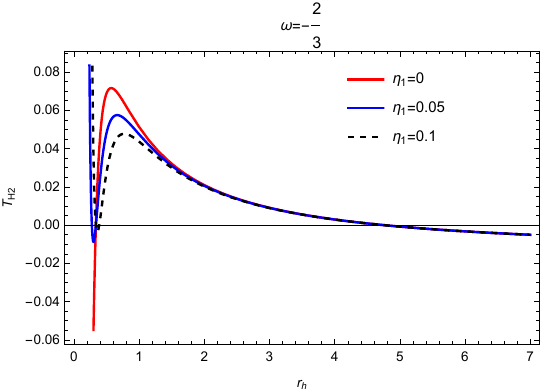}
\caption{}
\label{fig:graph2}
\end{subfigure}

\vspace{0.3cm}

\begin{subfigure}{\columnwidth}
\centering
\includegraphics[width=\linewidth]{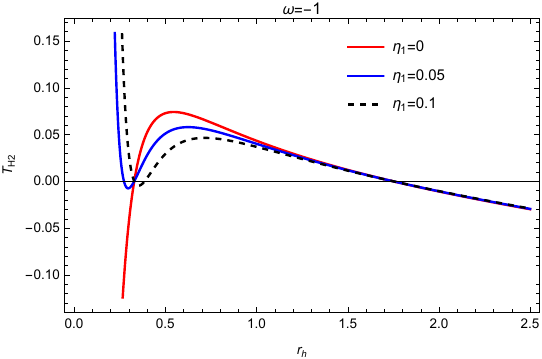}
\caption{}
\label{fig:graph3}
\end{subfigure}

\caption{Temperature ($ T_{H2} $) vs radius of event horizon ($ r_{h} $) graph for $a=0.1$  , $ \alpha=0.1 $, $ b=0.05 $, $ Q=0.3 $ and $ l_{p}=1 $.}
\label{Tem2}

\end{figure}

Based on the Bekenstein approach \cite{5}, when a particle with energy $E$ and size $R$ is absorbed by the black hole, there is a minimal increase in its area as
\begin{eqnarray}\label{eq.13}
(\Delta A)_{min}\geq 4 (\ln 2)l^{2}_{p}E R.
\end{eqnarray}
At the quantum mechanical level, the size of the particle can be described in terms of the uncertainty in its position, i.e. $ R \simeq \delta x \simeq r_{h} $. Then, Eq. \eqref{eq.13} can be rewritten as 
\begin{eqnarray}
(\Delta A)_{min}\geq 4 (\ln 2)l^{2}_{p}E \delta x.
\end{eqnarray}
Using Eqs. \eqref{eq.5} and \eqref{eq.6}, we get
\begin{eqnarray}\label{eq.15}
(\Delta A_{1})_{min}\geq 4 (\ln 2)l^{2}_{p} \left(1-\frac{\eta_{1}l_{p}}{\delta x} \right)
\end{eqnarray}
and
\begin{eqnarray}\label{eq.16}
(\Delta A_{2})_{min}\geq 4 (\ln 2)l^{2}_{p}\left(1-\frac{3\eta_{2}l^{2}_{p}}{2(\delta x)^{2}} \right).
\end{eqnarray}
Establishing that the minimal increase in its entropy is $(\Delta S)_{min} = \ln(2)$, we can express Eqs. \eqref{eq.15} and \eqref{eq.16} as
\begin{eqnarray}\label{eq.17}
\frac{dA_{1}}{dS_{1}}\simeq \frac{(\Delta A_{1})_{min}}{(\Delta S_{1})_{min}}\simeq 4 l^{2}_{p} \left(1-\frac{\eta_{1}l_{p}}{\delta x} \right)
\end{eqnarray}
and 
\begin{eqnarray}\label{eq.18}
\frac{dA_{2}}{dS_{2}}\simeq \frac{(\Delta A_{2})_{min}}{(\Delta S_{2})_{min}}\simeq 4 l^{2}_{p}\left(1-\frac{3\eta_{2}l^{2}_{p}}{2(\delta x)^{2}} \right).
\end{eqnarray}
Defining the Hawking temperature as \cite{68}
\begin{eqnarray}
T_{H}=\frac{\kappa}{8\pi}\times \frac{dA}{dS},
\end{eqnarray} 
we can get the Hawking temperature corrected by the MDR given in Eqs. \eqref{1} and \eqref{2} as follows
\begin{eqnarray}\label{eq.20}
T_{H1}= \frac{r_{h}-M-br_{h}-\frac{r_{h}^{-3\omega}\alpha (1-3\omega)}{2}}{2\pi (r_{h}^{2}+a^{2})} \left(1-\frac{\eta_{1}l_{p}}{\delta x} \right)l^{2}_{p}
\end{eqnarray}
and
\begin{eqnarray}\label{eq.21}
T_{H2}&=& \frac{r_{h}-M-br_{h}-\frac{r_{h}^{-3\omega}\alpha (1-3\omega)}{2}}{2\pi (r_{h}^{2}+a^{2})}\cr&& \times\left(1-\frac{3\eta_{2}l^{2}_{p}}{2(\delta x)^{2}} \right)l^{2}_{p} .
\end{eqnarray}
According to Eqs. \eqref{eq.20} and \eqref{eq.21}, we show that the corrected Hawking temperature is influenced by the quantum correction parameters $ \eta_{i}(i=1,2) $ resulting from the modified dispersion relation (MDR), the quintessence parameter $ \alpha $, and the parameter related to the cloud string $ b $. If $ \eta_{i},b,\alpha \rightarrow 0 $, we can get the original Hawking temperature of the Kerr-Newman black hole. We examine the corrected Hawking temperatures $ T_{H1} $ and $ T_{H2} $ for the cases where $ \omega=-\frac{1}{3} $, $ \omega=-\frac{2}{3} $, and $ \omega=-1 $, while varying the parameters $ \eta_{i} $. It is further acknowledged that the singularity occurs when $ r_{h} $ decreases because of the influence of MDR. The changes in $ T_{H1} $ and $ T_{H2} $ as a function of $ r_{h} $ are illustrated in Figs. \ref{Tem1} and \ref{Tem2}.

Then, the MDR-corrected entropy of the black hole can be studied by using Eqs. \eqref{eq.17} and \eqref{eq.18}.

From Eq. \eqref{eq.17}, we get
\begin{eqnarray}\label{eq.22}
S_{1}=S_{0}+\frac{2\pi \eta_{1}}{l_{p}}\sqrt{\frac{A}{4\pi}-a^{2}}.
\end{eqnarray}
Again, from Eq. \eqref{eq.18}, we get
\begin{eqnarray}\label{eq.23}
S_{2}=S_{0}+\frac{3 \pi \eta_{2}}{2}\log(A-4\pi a^{2}),
\end{eqnarray}
where $ S_{0}=\frac{A}{4l_{p}^{2}} $ is the original entropy of the Kerr-Newman black hole. Based on Eqs. \eqref{eq.22} and \eqref{eq.23}, it is clear that the corrected entropy is not affected by the quintessence but the MDR-correction parameters only. If $ \eta_{1}$ and $\eta_{2}$ tends to zero,  we arrive at the original entropy of the black hole. One notable observation from the expression is the presence of a logarithmic correction term in the case of $ S_{2} $, which is found in many quantum gravity theories, while $ S_{1} $ lacks such a term.

Next, we calculate the MDR-corrected heat capacity of the Kerr-Newman black hole surrounded by quintessence and a cloud of string using the relation
\begin{eqnarray}
C=\frac{dM}{dT}=\frac{\partial M}{\partial r}\frac{\partial r}{\partial T}.
\end{eqnarray}
Using Eqs. \eqref{eq.12} and \eqref{eq.20}, we get heat capacity of the black hole corrected by the MDR given in Eq. \eqref{1} as
\begin{eqnarray}\label{eq.25}
C_{1}&=&2\pi r_{h}(a^{2}+r_{h}^{2})^{2}\Big[  3r_{h}\alpha \omega -r_{h}^{3\omega}( a^{2}+Q^{2}\cr&& +(b-1)r_{h}^{2}) \Big]  \Big[r_{h}^{3\omega}l_{p}^{2} \lbrace  Q^{2}r_{h}^{2}(3r_{h}-4l_{p}\eta_{1})\cr&& +a^{4}(r_{h}-2l_{p}\eta_{1}) +(b-1)r_{h}^{4}(r_{h}-2l_{p}\eta_{1})\cr&& + a^{2}\lbrace Q^{2}(r_{h}-2l_{p}\eta_{1}) -r_{h}^{2}((b-4)r_{h}+4l_{p}\eta_{1})\rbrace\rbrace \cr&& +3r_{h}\alpha \omega l_{p}^{2} \lbrace -2r_{h}^{3} +a^{2}l_{p}\eta_{1} +3l_{p}r_{h}^{2}\eta_{1} -3(a^{2}+r_{h}^{2})\cr&& \times (r_{h}-l_{p}\eta_{1})\omega\rbrace \Big]^{-1}.
\end{eqnarray}
Again, using Eqs. \eqref{eq.12} and \eqref{eq.21}, we get
\begin{eqnarray}\label{eq.26}
C_{2}&=&4\pi r_{h}^{2}(a^{2}+r_{h}^{2})\left[3r_{h}\alpha \omega-r_{h}^{3\omega}(a^{2}+Q^{2}+(b-1)r_{h}^{2}) \right] \cr&& \times \Big[r_{h}^{3\omega}l_{p}^{2}\lbrace 2a^{2}\left( a^{2}+Q^{2}\right)r_{h}^{2}-2\left(a^{2}(b-4)-3Q^{2} \right)\cr&& \times r_{h}^{4}+2(b-1)r_{h}^{6} - 3l_{p}^{2}(3a^{2}(a^{2}+Q^{2})\cr&&+(a^{2}(b+4)+5Q^{2})r_{h}^{2}+3(b-1)r_{h}^{4})\eta_{2}) \rbrace \cr&& +3r_{h}\alpha \omega l_{p}^{2}( -4r_{h}^{4}+6l_{p}^{2}(a^{2}+2r_{h}^{2})\eta_{2}-3(a^{2}+r_{h}^{2})\cr&& \times(2r_{h}^{2}-3\eta_{2}l_{p}^{2})\omega)  \Big]^{-1}.
\end{eqnarray}

\begin{table*}[]
\caption{Phase transition points of the heat capacity (values of $r_h$) for different values of the MDR parameter $\eta_{1}$ and  $\omega$.}
\label{tab 1}
\centering
\begin{tabular}{cccccccc}

\multicolumn{1}{c}{$\omega$} & \multicolumn{1}{c}{$\eta_{1}=0$} & \multicolumn{1}{c}{$\eta_{1}=0.05$} & \multicolumn{1}{c}{$\eta_{1}=0.1$} \\ \hline
\multirow{1}{*}{-1/3} & 0.605139  &  0.0761471, 0.623871 &   0.13609, 0.649347  \\ \hline

\multirow{1}{*}{-2/3} &  0.570672 & 0.0768964, 0.587156 & 0.136839, 0.607387 \\ \hline                

\multirow{1}{*}{-1} &  0.545197 & 0.0768964, 0.559433 & 0.13609, 0.575917 \\ \hline
\end{tabular}
\end{table*}

\begin{table*}[]
\caption{Phase transition points of the heat capacity (values of $r_h$) for different values of the MDR parameter $\eta_{2}$ and  $\omega$.}
\label{tab 2}
\centering
\begin{tabular}{cccccccc}

\multicolumn{1}{c}{$\omega$} & \multicolumn{1}{c}{$\eta_{1}=0$} & \multicolumn{1}{c}{$\eta_{1}=0.05$} & \multicolumn{1}{c}{$\eta_{1}=0.1$} \\ \hline
\multirow{1}{*}{-1/3} & 0.604629  &  0.301163, 0.708647 &   0.362238, 0.828889  \\ \hline

\multirow{1}{*}{-2/3} &  0.570513 & 0.298539, 0.663557 & 0.357944, 0.769484 \\ \hline                

\multirow{1}{*}{-1} &  0.545463 & 0.297107, 0.625624 & 0.353649, 0.715805 \\ \hline
\end{tabular}
\end{table*}

\begin{figure}[h!]
\centering

\begin{subfigure}{\columnwidth}
\centering
\includegraphics[width=\linewidth]{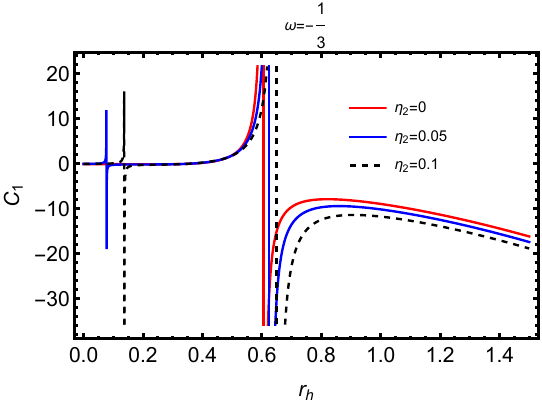}
\caption{}
\label{fig:graph1}
\end{subfigure}

\vspace{0.3cm}

\begin{subfigure}{\columnwidth}
\centering
\includegraphics[width=\linewidth]{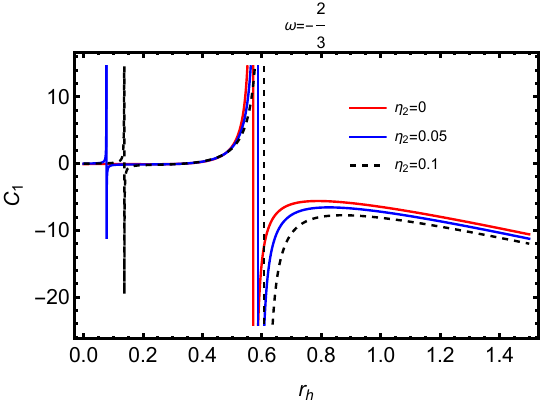}
\caption{}
\label{fig:graph2}
\end{subfigure}

\vspace{0.3cm}

\begin{subfigure}{\columnwidth}
\centering
\includegraphics[width=\linewidth]{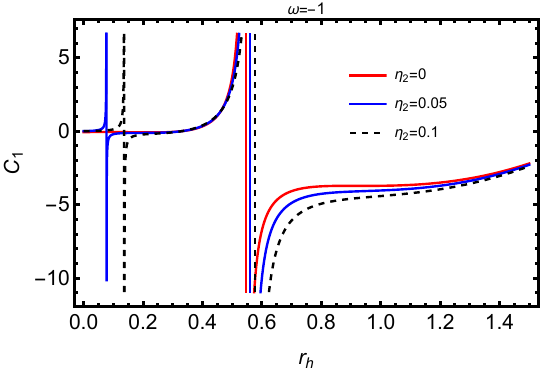}
\caption{}
\label{fig:graph3}
\end{subfigure}

\caption{Heat capacity ($ C_{1} $) vs radius of event horizon ($ r_{h} $) graph for $a=0.1$  , $ b=0.05 $, $ \alpha=0.1 $, $ Q=0.3 $ and $ l_{p}=1 $}
\label{Heat1}

\end{figure}

\begin{figure}[h!]
\centering

\begin{subfigure}{\columnwidth}
\centering
\includegraphics[width=\linewidth]{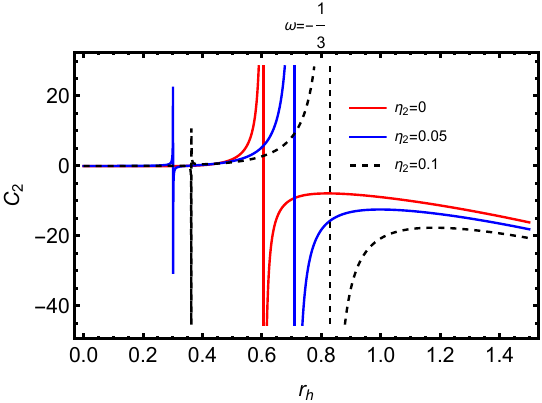}
\caption{}
\label{fig:graph1}
\end{subfigure}

\vspace{0.3cm}

\begin{subfigure}{\columnwidth}
\centering
\includegraphics[width=\linewidth]{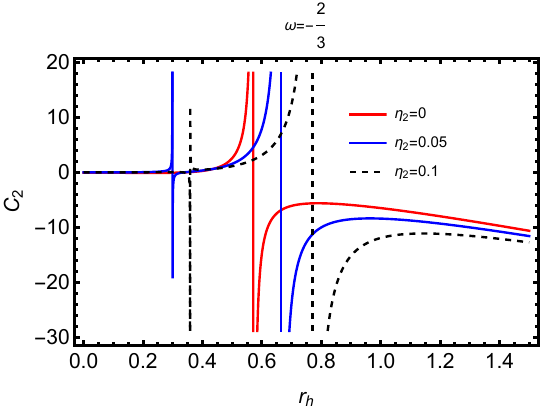}
\caption{}
\label{fig:graph2}
\end{subfigure}

\vspace{0.3cm}

\begin{subfigure}{\columnwidth}
\centering
\includegraphics[width=\linewidth]{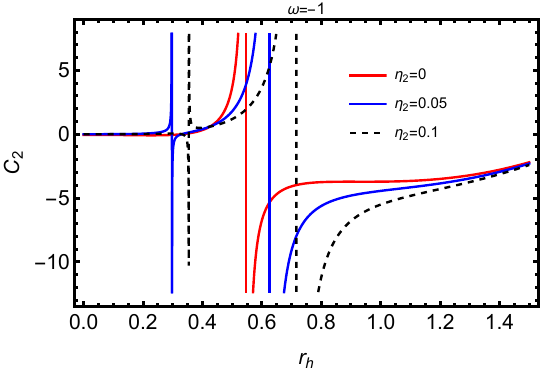}
\caption{}
\label{fig:graph3}
\end{subfigure}

\caption{Heat capacity ($ C_{2} $) vs radius of event horizon ($ r_{h} $) graph for $a=0.1$  , $ b=0.05 $, $ \alpha=0.1 $, $ Q=0.3 $ and $ l_{p}=1 $}
\label{Heat2}

\end{figure}

The graphs showing the changes in heat capacity $ C_{1}$ and $ C_{2} $ in relation to $ r_{h} $ are presented in Figs. \ref{Heat1} and \ref{Heat2}. The graphs clearly demonstrate that a dual phase transition occurs when we take into account the effects of MDR, whereas only a single phase transition is seen when MDR is not considered. The impact of MDR is small in case of $ C_{1} $ but a little bit larger in case of $ C_{2} $. The phase transition points of $ C_{1} $ and $ C_{2} $ for different values of $ \omega $ and $ \eta $ are shown in Tables \ref{tab 1} and \ref{tab 2} respectively.  It is observed that the position of the phase transition shifts to larger 
values of  $r_h$ as the MDR parameters $\eta_{1}$ and $\eta_{2}$ increase.
\begin{figure}[h!]
\centering

\begin{subfigure}{\columnwidth}
\centering
\includegraphics[width=\linewidth]{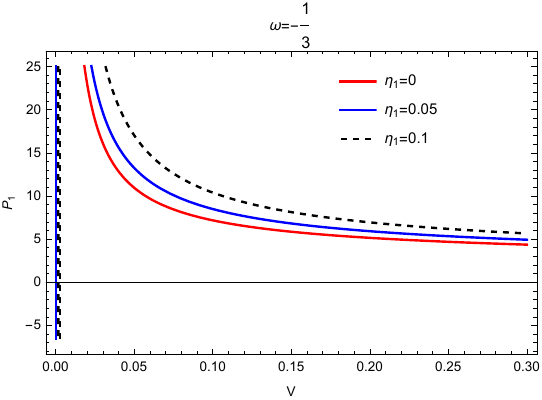}
\caption{}
\label{fig:graph1}
\end{subfigure}

\vspace{0.3cm}

\begin{subfigure}{\columnwidth}
\centering
\includegraphics[width=\linewidth]{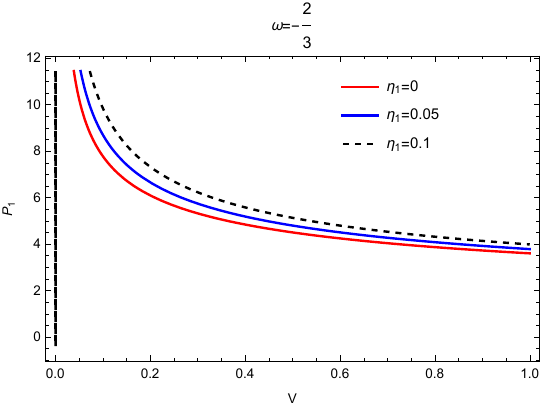}
\caption{}
\label{fig:graph2}
\end{subfigure}

\vspace{0.3cm}

\begin{subfigure}{\columnwidth}
\centering
\includegraphics[width=\linewidth]{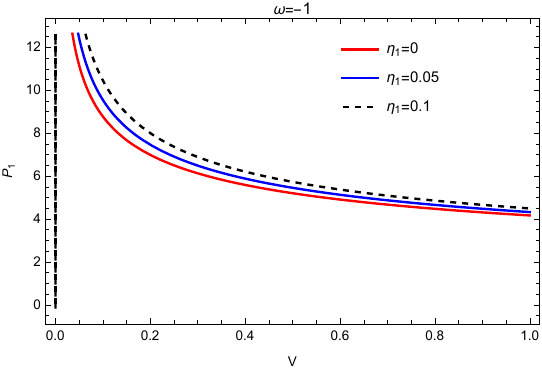}
\caption{}
\label{fig:graph3}
\end{subfigure}

\caption{Pressure $ P_{1} $ vs $ V $ graph for $a=0.1$  , $ \alpha=0.1 $, $ b=0.05 $, $ Q=0.3 $, $ T_{H1}=1 $ and $ l_{p}=1 $.}
\label{Pass1}

\end{figure}

\begin{figure}[h!]
\centering

\begin{subfigure}{\columnwidth}
\centering
\includegraphics[width=\linewidth]{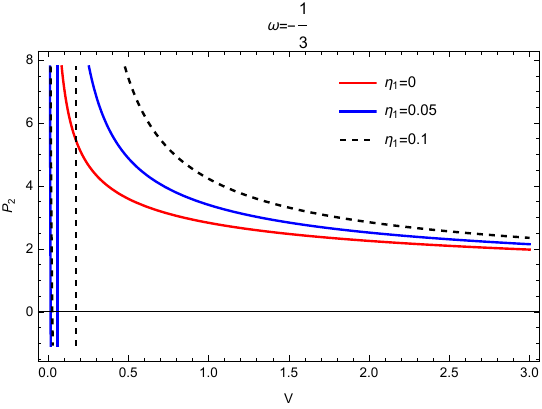}
\caption{}
\label{fig:graph1}
\end{subfigure}

\vspace{0.3cm}

\begin{subfigure}{\columnwidth}
\centering
\includegraphics[width=\linewidth]{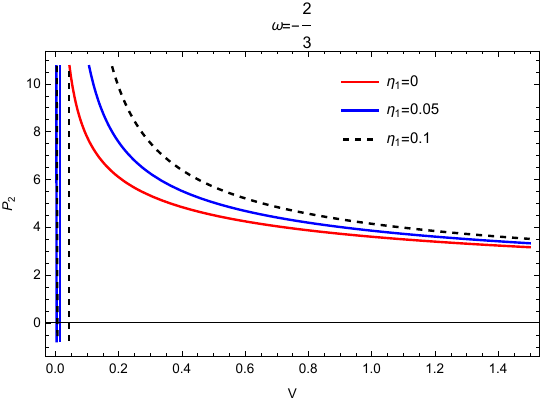}
\caption{}
\label{fig:graph2}
\end{subfigure}

\vspace{0.3cm}

\begin{subfigure}{\columnwidth}
\centering
\includegraphics[width=\linewidth]{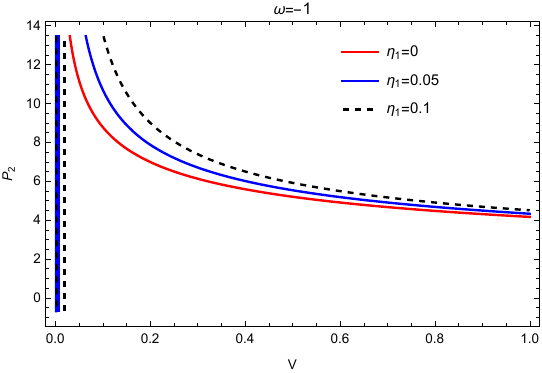}
\caption{}
\label{fig:graph3}
\end{subfigure}

\caption{Pressure $ P_{1} $ vs $ V $ graph for $a=0.1$  , $ \alpha=0.1 $, $ b=0.05 $, $ Q=0.3 $, $ T_{H1}=1 $ and $ l_{p}=1 $.}
\label{Pass2}

\end{figure}

The expression of the heat capacity given in Eqs. \eqref{eq.25} and \eqref{eq.26} is found to be zero when 
\begin{eqnarray}
3r_{h}\alpha \omega-r_{h}^{3\omega}\left( a^{2}+Q^{2}+(b-1)r_{h}^{2}\right) =0.
\end{eqnarray}
For $ \omega =-\frac{1}{3} $, the solutions are
\begin{eqnarray}\label{eq.28}
r_{h}=\pm \sqrt{\frac{a^{2}+Q^{2}}{1-b-\alpha}}.
\end{eqnarray}
For $ \omega=-\frac{2}{3} $, the only significant real solution is
\begin{eqnarray}\label{eq.29}
r_{h}&=&\frac{1-b-\frac{(1-b)^{2}}{\zeta}-\zeta}{6\alpha}.
\end{eqnarray}
For $ \omega =-1 $, there are four solutions
\begin{eqnarray}\label{eq.30}
r_{h}=\pm \sqrt{\frac{1-b\pm \sqrt{1-2b+b^{2}-12a^{2}\alpha - 12Q^{2}\alpha}}{6\alpha}},
\end{eqnarray}
where,\\ $ \zeta =\Big[(b-1)^{3}+54(a^{2}+Q^{2})\alpha^{2}+6\sqrt{3}\\ \sqrt{(a^{2}+Q^{2})\alpha^{2}\left( (b-1)^{3}+27(a^{2}+Q^{2})\alpha^{2}\right) }\Big]^{-\frac{1}{3}}. $
The vanishing of the heat capacity suggests the formation of a black hole remnant. For $ \omega=-\frac{1}{3} $, $ \omega=-\frac{2}{3} $ and $ \omega=-1 $, the heat capacity of the black hole vanishes for the values of $ r_{h} $ given in Eqs. \eqref{eq.28}, \eqref{eq.29} and \eqref{eq.30}. The mass of the black hole at which the heat capacity vanishes gives the remnant mass. From the above conditions, we can say that the quintessence and the cloud string parameters plays an important role in the formation of the black hole remnant while the MDR does not affect the formation of black hole remnant.

Next, we deduced the equation of state of the black hole by using the relation of pressure $ P $ and the quintessence parameter $ \alpha $ \cite{69},
\begin{eqnarray}
P=-\frac{3}{2}\frac{\alpha \omega^{2}}{r_{h}^{3(\omega+1)}}.
\end{eqnarray}
And the thermodynamic volume is defined as \cite{70}
\begin{eqnarray}\label{eq.32}
V=\frac{\partial M}{\partial P}=\frac{r_{h}^{3}}{3\omega^{2}}.
\end{eqnarray}
Using Eq. \eqref{eq.32}, we can express $ r_{h} $ in terms of $ V $ as
\begin{eqnarray}\label{eq.33}
r_{h}=\left( 3V\omega^{2} \right)^{\frac{1}{3}}. 
\end{eqnarray}
By using Eq. \eqref{eq.33} and solving $ \alpha $ from Eqs. \eqref{eq.20} and \eqref{eq.21}, we can express the equation of state effected by the two MDR as 
\begin{eqnarray}
P_{1}&=&\omega \Big[4\times 3^{1/3}\pi T_{H1}\left( V\omega^{2}\right) ^{1/3}\lbrace 3V\omega^{2}+3^{1/3}a^{2}\cr&&\times(V\omega^{2})^{1/3} \rbrace + l_{p}^{2}\lbrace 3\left( b-1\right) V\omega^{2}\cr&& +3^{1/3}\left( a^{2}+Q^{2}\right) \left( V\omega^{2}\right) ^{1/3}\rbrace \cr&& - l_{p}^{2}\eta_{1} \left\lbrace a^{2}+Q^{2}+3^{2/3}\left(b-1 \right)(V\omega^{2})^{2/3}  \right\rbrace \Big]\cr&& \times\Big[6\times 3^{1/3}l_{p}^{2}(V\omega^{2})^{4/3} \lbrace l_{p}\eta_{1} - 3^{1/3}\cr&&\times(V\omega^{2})^{1/3} \rbrace \Big]^{-1},
\end{eqnarray}
and
\begin{eqnarray}
P_{2}&=&\omega \Big[24 \pi T_{H2} V\omega^{2} \left(a^{2}+3^{2/3}(V\omega^{2})^{2/3} \right)-3l_{p}^{2}\eta_{2} \cr&& \times \left\lbrace a^{2}+Q^{2}+3^{2/3}\left(b-1 \right) (V\omega^{2})^{2/3} \right\rbrace \cr&& +2\times 3^{2/3}l_{p}^{2}(V\omega^{2})^{2/3} \lbrace a^{2}+Q^{2}\cr&& +3^{2/3}\left(b-1 \right)(V\omega^{2})^{2/3}\rbrace \Big]\Big[6\times 3^{1/3}l_{p}^{2}(V\omega^{2})^{4/3} \cr&& \times\left\lbrace 3l_{p}^{2}\eta_{2} -2\times 3^{2/3}(V\omega^{2})^{2/3}\right\rbrace \Big]^{-1}.
\end{eqnarray}
The isotherm graph of $ P$ vs $V $ for the two MDR with varying parameters $ \eta_{i} $ is illustrated in Figs. \ref{Pass1} and \ref{Pass2} for $ \omega=-\frac{1}{3} $, $ \omega=-\frac{2}{3} $, and $ \omega=-1 $. From the graph, it is clear that singularity is occur at the very small value of $ V $ due to the MDR-deformation parameter $ \eta_{i} $. A notable observation from the graph in Fig. \ref{Pass1} is that singularity is present only at small values of $ V $ when $ \eta_{1} \geq 1$ for $ \omega = -\frac{2}{3} $ and $ \omega = -1 $.

\section{Null geodesics of the Kerr-Newman black hole with quintessence and cloud of string}
In order to analyze the null geodesics equation within the space-time of the Kerr-Newman black hole surrounded by quintessence and cloud of strings, we utilize the Hamilton-Jacobi equation,
\begin{eqnarray}\label{eq.36}
\frac{\partial S}{\partial \lambda}=-\frac{1}{2}g^{\mu\nu}\frac{\partial S}{\partial x^{\mu}}\frac{\partial S}{\partial x^{\nu}},
\end{eqnarray}
where $ S $ is the Jacobi action and $ \lambda $ is an affine parameter of the curve.

To study the Hamilton-Jacobi equation, we carry out the separation of variable on the Jacobi action as
\begin{eqnarray}\label{eq.37}
S=-Et+L\varphi +S_{r}(r)+S_{\theta}(\theta),
\end{eqnarray}
where $ S_{r}(r) $ and $ S_{\theta}(\theta) $ are the functions of the radial and angular coordinates respectively.

Using the conserved energy $ E $ and angular momentum $ L $ and substituting Eq. \eqref{eq.37} into Eq. \eqref{eq.36}, we obtain 
\begin{eqnarray}
\left(\frac{d S_{\theta}}{d \theta} \right)^{2}+E^{2}a^{2}\sin^{2}\theta -E^{2}a^{2}+L^{2}\cot^{2}\theta =\cr-\Delta\left(\frac{d S_{r}}{dr} \right)^{2}+(r^{2}+a^{2})^{2}E^{2} -\frac{2E(r^{2}+a^{2})aL}{\Delta}\cr +\frac{a^{2}L^{2}}{\Delta}-E^{2}a^{2}+2aEL-L^{2}.
\end{eqnarray}
By utilizing Carter's constant $ K $, we can divide the radial component from the angular component.
\begin{eqnarray}\label{eq.39}
\Delta\left(\frac{d S_{r}}{dr}\right)^{2}-(r^{2}+a^{2})^{2}E^{2} +\frac{2E(r^{2}+a^{2})aL}{\Delta}-\frac{a^{2}L^{2}}{\Delta}\cr +E^{2}a^{2}-2aEL +L^{2}=-K
\end{eqnarray}
and
\begin{eqnarray}\label{eq.40}
\left(\frac{d S_{\theta}}{d \theta} \right)^{2}+E^{2}a^{2}\sin^{2}\theta -E^{2}a^{2}+L^{2}\cot^{2}\theta=K.
\end{eqnarray}
Using the relations of canonically conjugate momentum,
\begin{eqnarray}\label{eq.41}
\frac{\partial S_{r}}{\partial r}=p_{r}=\frac{\Sigma}{\Delta}\frac{dr}{d\lambda},
\end{eqnarray}
and
\begin{eqnarray}\label{eq.42}
\frac{\partial S_{\theta}}{\partial \theta}=p_{\theta}=\Sigma\frac{d\theta}{d\lambda}.
\end{eqnarray}
Inserting Eq. \eqref{eq.41} and Eq. \eqref{eq.42} into Eq. \eqref{eq.39} and Eq. \eqref{eq.40} respectively, we have
\begin{eqnarray}\label{eq.43}
\Sigma\frac{dr}{d\lambda}=\pm\sqrt{R}
\end{eqnarray}
and
\begin{eqnarray}\label{eq.44}
\Sigma \frac{d\theta}{d\lambda}=\pm \sqrt{\Theta},
\end{eqnarray}
where 
\begin{eqnarray}
R=\left[E(r^{2}+a^{2})-aL \right]^{2}-\Delta\left[K+(L-Ea)^{2} \right],  
\end{eqnarray}
and
\begin{eqnarray}
\Theta= K+\cos^{2}\theta \left(E^{2}a^{2}-\frac{L^{2}}{\sin^{2}\theta} \right). 
\end{eqnarray}
Then we can write the Jacobi action as
\begin{eqnarray}\label{eq.47}
S=-Et+L\varphi + \int^{\theta} \Theta\,d\theta+\int^{r}\Delta^{-1} R\,dr.
\end{eqnarray}
Differentiating Eq. \eqref{eq.47} with respect to $ E $ and $ L $, we obtain
\begin{eqnarray}\label{eq.48}
t&=&\int^{\theta}\frac{2Ea^{2}\cos^{2}\theta}{2\sqrt{\Theta}}\,d\theta +\int^{r}\frac{\Delta^{-1}}{\sqrt{R}}\Big[\left(E(r^{2}+a^{2})-aL \right)\cr&& \times(r^{2}+a^{2}) +\Delta a(L-Ea) \Big]\,dr
\end{eqnarray}
and
\begin{eqnarray}\label{eq.49}
\varphi &=&\int^{\theta}\frac{L\cot^{2}\theta}{\sqrt{\Theta}}\,d\theta +\int^{r}\frac{\Delta^{-1}}{\sqrt{R}}\Big[\left(E(r^{2}+a^{2})-aL \right)a\cr&& +\Delta(L-Ea) \Big]\,dr.
\end{eqnarray}
Taking derivative with respect to the affine parameter $ \lambda $ in Eqs. \eqref{eq.48} and \eqref{eq.49}, we obtain
\begin{eqnarray}\label{eq.50}
\Sigma \frac{d t}{d \lambda}&=&aL-Ea^{2}\sin^{2}\theta +\frac{1}{\Delta}\Big[E(r^{2}+a^{2})^{2}\cr&& -aL(r^{2}+a^{2})\Big]
\end{eqnarray}
and
\begin{eqnarray}\label{eq.51}
\Sigma \frac{d\varphi}{d\lambda}=\frac{L}{\sin^{2}\theta}-Ea+\frac{1}{\Delta}\Big[Ea(r^{2}+a^{2})-a^{2}L\Big].
\end{eqnarray}
Eqs. \eqref{eq.43}, \eqref{eq.44}, \eqref{eq.50} and \eqref{eq.51} represent the null geodesic equations of motion of the photon around the Kerr-Newman black hole surrounded by quintessence and cloud of string.

We now focus on the radial equation in which we get the effective potential as
\begin{eqnarray}
\left(\frac{dr}{d\lambda} \right)^{2}=V_{eff}, 
\end{eqnarray}
where
\begin{eqnarray}\label{eqn_Veff}
V_{eff}&=&\frac{1}{r^{4}}\Big[\lbrace E(r^{2}+a^{2})-aL \rbrace^{2}-\Delta\lbrace K\cr&& +(L-Ea)^{2} \rbrace\Big].
\end{eqnarray}
\begin{figure}[htbp]
\centering
\includegraphics[width=\columnwidth]{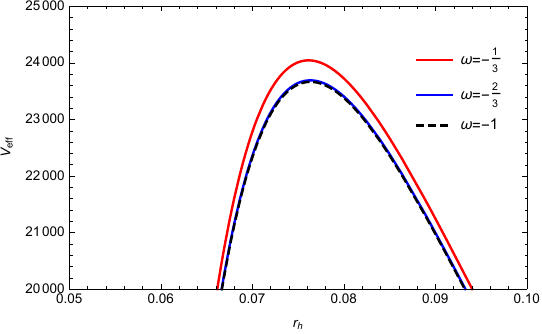}
\caption{Variation of the effective potential with respect to $r$ for $ \omega=-1/3 $, $ \omega=-2/3 $, $ \omega=-1 $.}
\label{Veff1}
\end{figure}

\begin{figure}[h!]
\centering

\begin{subfigure}{\columnwidth}
\centering
\includegraphics[width=\linewidth]{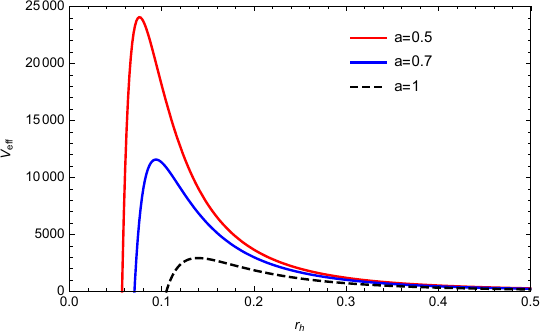}
\caption{}
\label{fig:graph1}
\end{subfigure}

\vspace{0.3cm}

\begin{subfigure}{\columnwidth}
\centering
\includegraphics[width=\linewidth]{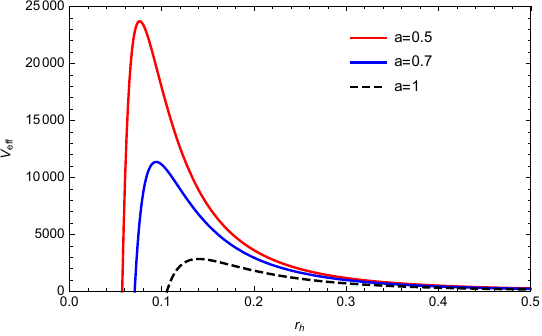}
\caption{}
\label{fig:graph2}
\end{subfigure}

\vspace{0.3cm}

\begin{subfigure}{\columnwidth}
\centering
\includegraphics[width=\linewidth]{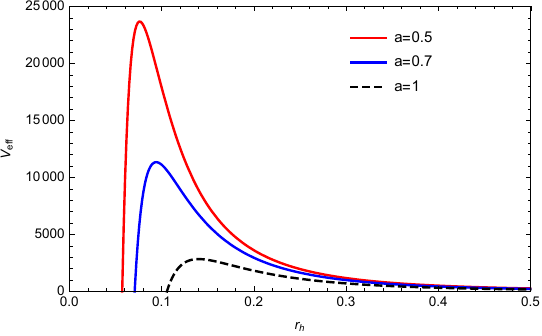}
\caption{}
\label{fig:graph3}
\end{subfigure}

\caption{(a)Variation of the effective potential with respect to $ r $ for $ \omega=-1/3 $ at different values of $ a $. (b)Variation of the effective potential with respect to $ r $ for $ \omega=-2/3 $ at different values of $ a $. (c)Variation of the effective potential with respect to $ r $ for $ \omega=-1 $ at different values of $ a $.}
\label{Veff2}

\end{figure}

\begin{figure}[h!]
\centering

\begin{subfigure}{\columnwidth}
\centering
\includegraphics[width=\linewidth]{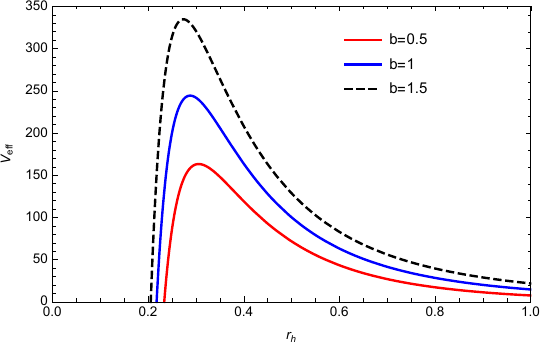}
\caption{}
\label{fig:graph1}
\end{subfigure}

\vspace{0.3cm}

\begin{subfigure}{\columnwidth}
\centering
\includegraphics[width=\linewidth]{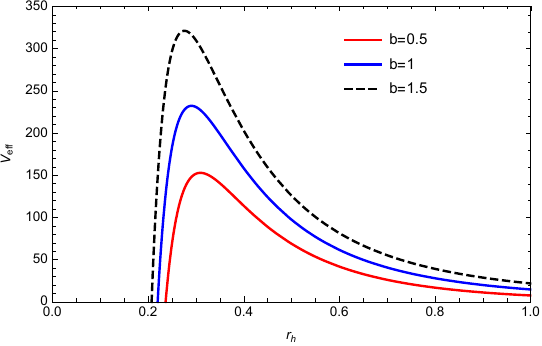}
\caption{}
\label{fig:graph2}
\end{subfigure}

\vspace{0.3cm}

\begin{subfigure}{\columnwidth}
\centering
\includegraphics[width=\linewidth]{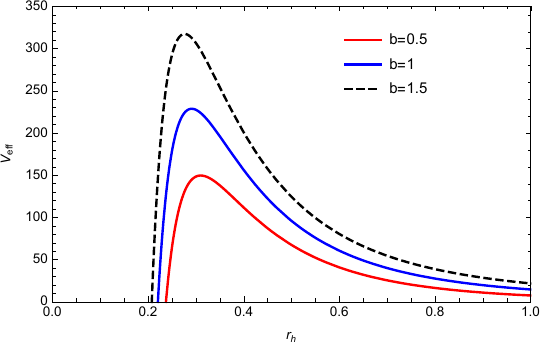}
\caption{}
\label{fig:graph3}
\end{subfigure}

\caption{(a)Variation of the effective potential with respect to $ r $ for $ \omega=-1/3 $ at different values of $ b $. (b)Variation of the effective potential with respect to $ r $ for $ \omega=-2/3 $ at different values of $ b $. (c)Variation of the effective potential with respect to $ r $ for $ \omega=-1 $ at different values of $ b $.}
\label{Veff3}

\end{figure}
Fig. \ref{Veff1} illustrates the variation of the effective potential $V_{eff}$ with respect to $r$ for different values of the $\omega=-1/3,-2/3,-1$ while keeping other parameters fixed. The effective potential exhibits a typical barrier-like structure with a single peak  corresponding to the unstable circular photon orbit. The peak of the effective potential slightly lowers and shifts marginally in the radial direction as $\omega$ decreases from $-1/3$ to -1. This indicates that that increasing the dark-energy–like contribution  effectively reduces the strength of the gravitational trapping near the photon sphere and also modifies the location of the unstable null geodesics. 

In Fig. \ref{Veff2} , we illustrates the effect of  the parameter $a$ on the effective potential. Unlike the moderate influence of $\omega$, increasing $a$ noticeably decreases the height of the potential peak height and slightly shifts the maximum toward larger radii. This indicates that $a$ weakens the gravitational trapping of null particles in the vicinity of the circular orbit. Fig. \ref{Veff3} shows the variation of the effective potential for different values of the could of string parameter $b$. In all cases, increasing $b$ significantly raises the height of the potential barrier and  shifts the peak toward smaller values of $r$. This indicates that $b$ strengthens the gravitational confinement of photons and enhances the instability of the circular null orbit.

Overall, while the qualitative behavior of the effective potential remains unchanged, its quantitative characteristics—such as the height and radial position of the maximum—are sensitive to the model parameters. These results clearly demonstrate how the underlying spacetime parameters control the stability and dynamics of circular null motion.

\section{Lyapunov exponent and instability of circular photon orbits}
We now investigate the instability of circular null geodesics through the Lyapunov exponent. We restrict the motion to the equatorial plane in which the Carter's constant $K$ must vanish ($K=0$) to ensure the photon does not deviate from the plane. Consequently, the effective potential in Eq. \eqref{eqn_Veff} becomes 
\begin{align}\label{eqn:veffnew}
V_{eff}=\frac{1}{r^{4}}\Big[\left[E(r^{2}+a^{2})-aL \right]^{2}-\Delta (L-Ea)^{2} \Big].
\end{align}

The radius of the circular photon orbit $r_c$ is obtained from the conditions \cite{71,72}
\begin{align}
V_{eff}(r_c)=0 \quad  \text{and }  \quad 
V'_{eff}\Big|_{r=r_c}=0.
\end{align}
\begin{figure}[h!]
\centering

\begin{subfigure}{\columnwidth}
\centering
\includegraphics[width=\linewidth]{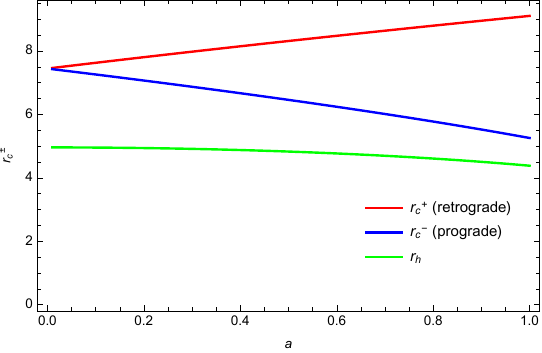}
\caption{}
\label{fig:rca}
\end{subfigure}

\vspace{0.3cm}

\begin{subfigure}{\columnwidth}
\centering
\includegraphics[width=\linewidth]{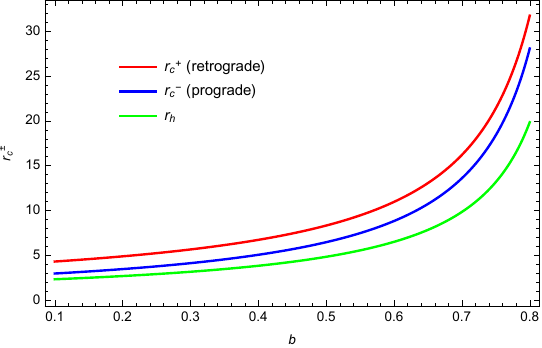}
\caption{}
\label{fig:rcb}
\end{subfigure}

\caption{(a)Variation of circular photon orbit radius $r_c^{\pm}$ for $ \omega=-1/3 $  with respect to (a) spin parameter $ a $ and (b) could of string parameter $b$.}
\label{fig:rc}

\end{figure}
Solving the circular photon orbit condition yields two branches corresponding to the prograde and retrograde photon orbits. The radii of the prograde and retrograde circular photon orbits are determined from
\begin{align}
\left(r_c^{\pm}\,\Delta'(r_c^{\pm})-4\Delta(r_c^{\pm})\right)^{2}-16a^{2}\Delta(r_c^{\pm})=0 ,
\end{align}
where the ``$+$'' sign corresponds to the retrograde photon orbit  and the ``$-$'' sign corresponds to the prograde photon orbit. 
The corresponding critical impact parameters are obtained as
\begin{align}\label{eqn:bc}
b_c^{\pm} =\dfrac{L}{E}= a+ \dfrac{r_c^{\pm}}{a\pm\sqrt{\Delta(r_c^{\pm})}}.
\end{align}
The circular photon orbits must lie outside the event horizon. As illustrated in Figs. \ref{fig:rca} and \ref{fig:rcb}, both the prograde  and retrograde  photon orbit radii lie outside  the event horizon radius. Hence, these photon trajectories occur in the exterior region of the black hole and represent physically admissible circular null geodesics.
The prograde photon orbit lies closer to the black hole horizon due to frame dragging, whereas the retrograde orbit occurs at a larger radius.

The variation of the photon circular orbit with the rotation parameter $a$ is shown in Fig. \ref{fig:rca}.  As the spin parameter increases, the retrograde photon orbit radius $r_c^{+}$ increases, whereas the prograde photon orbit radius $r_c^{-}$ gradually decreases. Photons moving in the same direction as the black hole rotation are dragged along by the rotating spacetime, enabling the prograde orbit to occur closer to the black hole. In contrast, photons moving opposite to the rotation experience an effective resistance from the dragging of inertial frames thereby shifting the retrograde orbit to larger radius. Fig. \ref{fig:rcb} shows the effect of the cloud of strings parameter $b$ on the photon circular orbits. It is observed that both $r_c^{+}$ and $r_c^{-}$ increase monotonically with increasing $b$. In contrast to the spin parameter, the cloud of strings does not introduce directional effects, and therefore both the prograde and retrograde photon orbits exhibit a similar increasing trend.

The instability of the circular photon orbit can be characterized by the Lyapunov exponent, which  is given by \cite{73}
\begin{align}
\lambda_{\pm} = \sqrt{\frac{V''_{eff}(r_c^{\pm})}{2\dot{t}^{2}}},
\end{align}
where $V''_{eff}$ denotes the second derivative of the effective potential evaluated at the circular photon orbit and $\dot{t}$ is obtained from the corresponding geodesic equation. A larger value of the Lyapunov exponent indicates that nearby photon trajectories diverge more rapidly from the circular orbit, implying a higher degree of orbital instability. Therefore, the Lyapunov exponent provides a quantitative measure of the instability timescale of the circular photon orbit. Using Eq. \eqref{eqn:veffnew} and evaluating it at the prograde/retrograde circular photon orbit, the corresponding Lyapunov exponent is obtained as
\begin{strip}
\begin{equation}
\lambda_\pm =
\sqrt{
\frac{
-\Biggl[
8\,r^2 + \dfrac{16\,r\,\Delta(r_c^{\pm})}{\Delta'(r_c^{\pm})} - \dfrac{16\,(r_c^{\pm})^2\,\Delta(r_c^{\pm})\,\Delta''(r_c^{\pm})}{\bigl(\Delta'(r_c^{\pm})\bigr)^2}
\Biggr]
}{
2 \,\Biggl[
a\,b_c(r_c^{\pm}) - a^2 + \dfrac{((r_c^{\pm})^2 + a^2)^2 - a\,b_c(r_c^{\pm})\,((r_c^{\pm})^2 + a^2)}{\Delta(r_c^{\pm})}
\Biggr]
}
}.
\end{equation}
\end{strip}
Here $\lambda_{\pm}$ represent the Lyapunov exponent associated with the retrograde $(+)$ and prograde $(-)$ circular photon orbits, respectively.
\begin{figure}[h!]
\centering

\begin{subfigure}{\columnwidth}
\centering
\includegraphics[width=\linewidth]{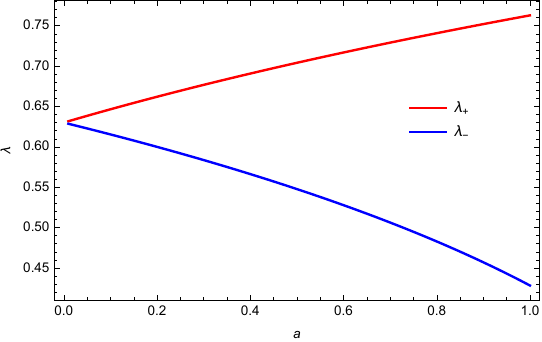}
\caption{}
\label{fig:lambdaa}
\end{subfigure}

\vspace{0.3cm}

\begin{subfigure}{\columnwidth}
\centering
\includegraphics[width=\linewidth]{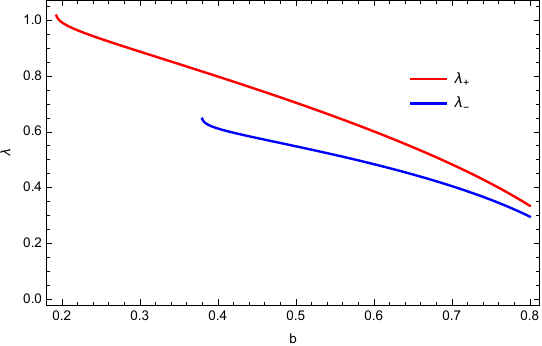}
\caption{}
\label{fig:lambdab}
\end{subfigure}

\caption{(a)Variation of circular photon orbit radius $r_c^{\pm}$ for $ \omega=-1/3 $  with respect to (a) spin parameter $ a $ and (b) could of string parameter $b$.}
\label{fig:lambda}

\end{figure}

The behavior of the Lyapunov exponent for the circular photon orbits is illustrated in Fig. \ref{fig:lambda}. Fig.  \ref{fig:lambdaa} shows the variation of the Lyapunov exponent $\lambda_{\pm}$ with respect to the spin parameter $a$. It is observed that the Lyapunov exponent corresponding to the retrograde orbit $\lambda_{+}$ increases with increasing $a$, whereas the Lyapunov exponent associated with the prograde orbit $\lambda_{-}$ decreases. This indicates that the instability of the retrograde photon orbit becomes stronger with increasing rotation of the black hole, while the prograde orbit becomes comparatively less unstable.

Fig.  \ref{fig:lambdab} depicts the variation of $\lambda_{\pm}$ with the cloud of strings parameter $b$. It is found that both $\lambda_{+}$ and $\lambda_{-}$ decrease monotonically with increasing $b$. This behavior suggests that the presence of the cloud of strings tends to suppress the instability of the circular photon orbits.
\section{Conclusion}
In this study, we explore how a modified dispersion relation influences the thermodynamic properties of the Kerr-Newman black hole in the presence of quintessence and a cloud of string. Initially, we compute the modified Hawking temperature of the black hole, taking into account two modified energy-momentum dispersion relations presented in Eqs. \eqref{1} and \eqref{2}. The modified Hawking temperature is impacted by quantum correction parameters $( \eta_{i} )$, which arise from the modified dispersion relation, the quintessence parameter \( \alpha \), and the cloud string parameter $( b )$. The graphs displayed in Figs. \ref{Tem1} and \ref{Tem2} clearly indicate that a singularity emerges as $( r_{h} )$ diminishes, a consequence of the modified dispersion relation. Subsequently, we derived the MDR-corrected entropy of the black hole. The corrected entropy is unaffected by quintessence but does depend on the MDR correction parameter $( \eta_{i} )$. A logarithmic correction term is present in the case of $( S_{2} )$, but absent in $( S_{1} )$. Following this, we calculated the MDR-corrected heat capacity of the black hole. The graph plotting heat capacity against $( r_{h} )$ reveals a dual phase transition when considering the effects of MDR, while a single phase transition is observed when these effects are disregarded. The vanishing of heat capacity implies the formation of a black hole remnant. The possible values of $ r_{h} $ where the remnant can be formed are given in Eqs. \eqref{eq.28}, \eqref{eq.29}, and \eqref{eq.30} for $ \omega=-\frac{1}{3} $, $ \omega=-\frac{2}{3} $, and $ \omega=-1 $. The quintessence and cloud string parameters affected the formation of the black hole remnant, whereas the MDR did not. Then, we analyse the equation of state concerning the two MDRs of the black hole. Based on the $ P $ vs $ V$ isotherm graph, it can be inferred that a singularity arises at a minimum value of $ V $ due to the MDR-deformation parameter $ \eta_{i} $. Lastly, we obtain the null-geodesic equations for the Kerr-Newman black hole in the presence of quintessence and a string cloud by utilizing the Hamilton-Jacobi equation and the corresponding effective potential governing null trajectories is obtained. The behavior of the effective potential reveals a typical potential barrier structure associated with unstable circular photon orbits. Our analysis shows that the decreasing the value of $\omega$ lowers the peak of the potential and modifies the location of the unstable photon orbit. Increasing the rotation parameter significantly decreases the height of the potential barrier and shifts the maximum toward larger radii, indicating a weakening of the gravitational trapping of photons. In contrast, increasing the cloud of strings parameter elevates the potential barrier and shifts the peak towards lower radius, thereby strengthening the confinement of photon trajectories around the black hole.
Furthermore, the radii of the prograde and retrograde circular photon orbits are determined. It is observed that the prograde orbit moves closer to the event horizon with increasing spin parameter due to frame dragging, while the retrograde orbit shifts to larger radii. In contrast, the cloud of strings parameter causes both prograde and retrograde photon orbit radii to increase monotonically, indicating that the surrounding string cloud modifies the spacetime structure without introducing directional effects.
Finally, the instability of the circular photon orbits is characterized through the Lyapunov exponent. The results show that the Lyapunov exponent corresponding to the retrograde orbit increases with increasing the the spin parameter while reducing that of the prograde orbit, revealing distinct stability behaviors induced by the rotation of the black hole. Moreover, increasing the cloud of strings parameter reduces the Lyapunov exponent for both prograde and retrograde orbits, indicating a suppression of orbital instability. These results highlight that the surrounding fields and rotation significantly influence the dynamics and stability of photon trajectories around the Kerr-Newman black hole surrounded by quintessence.

Overall, our results demonstrate that the modified dispersion relation, together with the effects of quintessence and the cloud of strings, plays an important role in modifying both the thermodynamic properties and the photon dynamics of the Kerr-Newman black hole surrounded by quintessence. The combined analysis of thermodynamic behavior and null geodesic structure provides deeper insight into the physical properties of black holes in the presence of quantum corrections and surrounding fields.

\section*{Acknowledgements}
The first author is being supported by the INSPIRE Fellowships of the Department of Science and Technology(DST),-New Delhi, India (INSPIRE Code: IF200576). 

\section*{Declaration of competing interest}  The authors declare that they have no known competing financial interests or personal relationships that could have appeared to influence the work reported in this paper

\end{document}